\documentclass[12pt,a4paper]{article}
\usepackage[numbers]{natbib}
\usepackage{authblk}
\usepackage[margin=1in]{geometry}

\RequirePackage{amsthm,amsmath,amsfonts,amssymb}

\RequirePackage{graphicx}

\usepackage[noend]{algpseudocode}
\usepackage{algorithm}
\usepackage{algorithmicx}
\usepackage{color}
\usepackage{enumitem}
\usepackage{graphics}
\usepackage{pagecolor}
\usepackage{soul}
\usepackage{subfigure}

\graphicspath{ {plots/} }

\newcommand{\timerev}[1]{\protect\overleftarrow{X}_{#1}^{(2)}}
\newcommand{\dd}{\,\mathrm{d}}

\newcommand{\algoref}[1]{Algorithm \ref{#1}}

\newcommand{\figref}[1]{Figure \ref{#1}}

\newcommand{\secref}[1]{Section \ref{#1}}

\providecommand{\keywords}[1]
{
  \small	
  \textbf{\textit{Keywords---}} #1
}

\title{Simulating bridges using confluent diffusions}

\author[1,2,3]{Paul A. Jenkins}
\author[3,4]{Murray Pollock\footnote{Corresponding author. Email: murray.pollock@newcastle.ac.uk}}
\author[1,4]{Gareth O. Roberts}
\author[5]{Michael S{\o}rensen}
\affil[1]{\small{Department of Statistics, University of Warwick, Coventry, CV4 7AL.}}
\affil[2]{Department of Statistics, University of Warwick, Coventry, United Kingdom, CV4 7AL.}
\affil[3]{The Alan Turing Institute, British Library, London, United Kingdom, NW1 2DB.}
\affil[4]{School of Mathematics, Statistics and Physics, Newcastle University, Newcastle-upon-Tyne, United Kingdom, NE1 7RU.}
\affil[5]{Department of Mathematical Sciences, University of Copenhagen, Universitetsparken 5, DK-2100 Copenhagen {\O}, Denmark.}

\begin{document}

\maketitle


\begin{abstract}
Diffusions are a fundamental class of models in many fields, including finance, engineering, and biology. Simulating diffusions is challenging as their sample paths are infinite-dimensional and their transition functions are typically intractable. In statistical settings such as parameter inference for discretely observed diffusions, we require simulation techniques for diffusions conditioned on hitting a given endpoint, which introduces further complication. In this paper we introduce a Markov chain Monte Carlo algorithm for simulating bridges of ergodic diffusions which (i) is \emph{exact} in the sense that there is no discretisation error, (ii) has computational cost that is linear in the duration of the bridges, and (iii) provides bounds on local maxima and minima of the simulated trajectory.  Our approach works directly on diffusion path space, by constructing a proposal (which we term a \emph{confluence}) that is then corrected with an accept/reject step in a pseudo-marginal algorithm. Our method requires only the simulation of unconditioned diffusion sample paths. We apply our approach to the simulation of Langevin diffusion bridges, a practical problem arising naturally in many situations, such as statistical inference in distributed settings.
\end{abstract}




\section{Introduction} \label{sec:intro}

Diffusions are popular within a variety of application areas as models for stochastic dynamical systems. These applications include, amongst others, the physical sciences \citep{sjs:pgd09}, the life sciences \citep{sc:gw06,csda:gw08}, and finance \citep{jfe:bs04,jpe:bs73,jof:ejp03, bjems:m73, jfe:m76}. Given a model for a dynamical system of interest it is natural to want to conduct inference on the model parameters. The complexity of inference for diffusions often necessitates the use of computational techniques such as Markov chain Monte Carlo (MCMC). These require sampling diffusion bridges with different parameter values, while ensuring the bridges sampled are coherent with the observed data.

To introduce diffusion bridges, first consider the following one-dimensional stochastic differential equation (SDE):
\begin{equation}\label{eq:sde_master}
\dd X_t = b(X_t) \dd t + \sigma(X_t)\dd W_t,\quad X_{t_0} = x_{t_0},\qquad t\in[t_0,T].
\end{equation}
We term an \emph{unconditioned diffusion} as being a weak solution to \eqref{eq:sde_master}, and if $X$ is conditioned on a terminal value, $X\mid X_T = x_T$, we term it a \emph{conditioned diffusion} or a \emph{diffusion bridge}.

Simulating diffusions is challenging. For very simple classes of diffusions in which the transition function is known, efficient schemes for simulating trajectories of unconditioned diffusions over long (temporal) durations ($T\gg 1$) are available---one can simply apply the strong Markov property and iteratively simulate partial trajectories. As such, the computational cost in this setting scales linearly in the number of iterations. However, in general the transition function for the majority of interesting classes of diffusions is intractable. For diffusion bridges the conditioning precludes the iterative chaining of partial trajectories, and so the scaling with the number of iterations is more complicated. 

Existing approaches for diffusion bridge simulation fall into two broad categories: methods based on simulating from time-discretisations (and thus approximations) of \eqref{eq:sde_master}, and those which rely on constructions directly on diffusion path-space for \eqref{eq:sde_master} (which we term \emph{path-space samplers}). The latter eliminates any discretisation error, but usually at a cost of introducing technical restrictions on $b$ and $\sigma$. In both cases various embedded Monte Carlo algorithms can be used (for instance, importance and rejection sampling or MCMC), with repeated simulation from an appropriately chosen proposal distribution. 

A key consideration is to ensure that the proposal and target bridge are well matched, while simultaneously accounting for the computational complexities of the proposal. One of the earliest breakthroughs was made by Pedersen \citep{sjs:p95}, who used a time-discretised realisation of the unconditioned diffusion \eqref{eq:sde_master} as a proposal. However, this approach suffers from degeneracy between the target and proposal distribution as increasingly fine time-discretisations are considered. Further work by \citep{spa:dh06, jbes:dg02, csda:gw08, ss:o92} attempt to directly resolve this, with other approaches such as `pulling' drifts \citep{spa:dh06}, guided and residual proposals \citep{b:smz17,arxiv:ms17,sc:wgbs17}, and the use of stochastic partial differential equations \citep{tsa:hsv09}, being similarly motivated. However, these approaches become increasingly computationally expensive with the fine resolutions required in settings where controlling or removing the bias is important, for instance in inference for partially observed diffusions \citep{bio:rs01}.

A conceptually novel alternative for simulating bridges of ergodic diffusions using a discretised approach was recently proposed by Bladt \& S\o{}rensen and subsequently extended \citep{jrssb:bfs16,b:bms20,b:bs14}. The proposal is formed by simulating only unconditioned diffusions---one whose start points coincides with that of the desired bridge, another whose start point coincides with the \emph{end point} of the bridge---these are spliced together at a well specified crossing time to form a single proposal path, which is then either accepted or rejected by means of a stochastic number of auxiliary unconditioned diffusions. The main strength of this approach is that only unconditioned diffusions are simulated, and so the desirable linear-in-time computational cost of such simulation is retained. This is in contrast to previously proposed approaches where the computational cost can grow exponentially in $T$. Termed the \emph{simple diffusion bridge} approach, this recent innovation forms one key ingredient of the work we introduce in this paper.

Alternatively, path-space rejection samplers allow sample paths of both unconditioned and conditioned diffusions to be simulated at finite collections of time points, without any approximation error. Following the initial work of \citep{aap:br05}, there has been considerable interest in developing unbiased estimators for classes of stochastic differential equations (including \citep{mcap:bpr08,aap:bcd17,mor:ch13,aap:js17,b:pjr16,or:rg15}). Methodologically these approaches follow a common form: sample paths are drawn from a target measure by means of drawing sample paths from an equivalent proposal measure in which access to the conditioned transition density is available, and are accepted with probability proportional to the Radon-Nikod{\'y}m derivative of the two measures. Path-space rejection sampling approaches have been applied to Bayesian inference for partially-observed diffusions \citep{b:bpr06,sd:brsv08,sjs:sprbf13}. However, a limitation is that the computational cost increases exponentially with bridge duration, and so are unsuitable in many applications.

It would be appealing to incorporate within the \emph{simple diffusion bridge} framework of Bladt \& S\o{}rensen the discretisation free path-space rejection samplers for unconditioned ergodic diffusions---this would reduce the computational cost of path-space rejection sampling approaches from exponential to linear with respect to bridge duration. However, the difficulty with a na\"ive embedding of these methodologies is it requires the identification of the joint crossing times of a collection of diffusions---both for the spliced proposal, \emph{and} the auxiliary trajectories used in determining acceptance. Instead, we resolve this issue by adapting ideas from \emph{$\epsilon$-strong} simulation within path-space rejection sampling \citep{beskos2012varepsilon,b:pjr16}. In particular, diffusion trajectories are simulated together with upper and lower convergent bounding processes which enfold (almost surely) the trajectory. We further develop methodology within this framework allowing us to unbiasedly estimate the probability that two such diffusion paths intersect. The resulting methodology, which we term the \emph{confluent diffusion bridge} approach, then allows us to overcome the challenge of embedding path-space rejection sampling within the simple diffusion bridge framework, and so attains the linear computational cost of the latter. We find practical application of our work motivated by \emph{Monte Carlo Fusion} \citep{jap:dpr19}, by considering the simulation of Langevin diffusion bridges in which a mixture of Gaussian distributions is the invariant measure of the unconditioned diffusion. 

We begin in \secref{sec:SDB} by outlining the simple diffusion bridge sampler of Bladt \& S\o{}rensen \cite{b:bs14}, and then in \secref{sec:PSRS} the pertinent aspects of path-space rejection sampling and $\varepsilon$-strong simulation. These methodologies together with our approach in \secref{sec:CDB} form the basis of the \emph{confluent diffusion bridge} (CDB) sampler introduced in this paper. In \secref{sec:numerical} we apply the CDB to numerical examples: in \secref{sec:ou_process}, we investigate the simulation of Ornstein--Uhlenbeck diffusion bridges with increasing duration, showing that the CDB sampler attains a linear computational scaling as opposed to the exponential scaling of competing methodologies; in \secref{sec:langevin_process} we consider the more complicated example of the Langevin diffusion bridge, which has a number of practical statistical applications. Finally, in \secref{sec:conclusion} we provide a brief discussion on applications and extensions of our work. 


\section{Simple diffusion bridges}\label{sec:SDB}

The \emph{simple diffusion bridge} sampler introduced by Bladt \& S\o{}rensen \cite{b:bs14}, obtains approximate draws from the conditioned diffusion bridge of \eqref{eq:sde_master}, with law which we denote $\mathbb{P}^{(x_0,x_T,T)}$. For clarity of exposition, we denote $\mathbb{P}^{(x_0,T)}$ as the unconditioned law. Broadly, the approach is an MCMC scheme in which a proposal bridge is obtained via rejection sampling, and then a Metropolis--Hastings correction is made to account for the discrepancy between the law of the proposal and the law of the target. The novelty of the method is the construction of the proposal \cite{b:bs14,b:bms20}, which we will now summarise. 

Bladt \& S\o{}rensen assume that a stationary distribution $\nu$ exists, and consider three independent diffusions $X^{(i)}$, $i=1,2,3$, solutions to \eqref{eq:sde_master}, started from $X_0^{(1)}=x_0$, $X_0^{(2)}=x_T$, and $X^{(3)}_{0}\sim \nu$, respectively. These are referred to as the \emph{forward}, \emph{backward}, and \emph{auxiliary diffusions} respectively. $\timerev{}$ is further defined as
\begin{equation*}
\timerev{t} = X^{(2)}_{T-t},\quad t\in[0,T],
\end{equation*}
and referred to as the \emph{time-reversed diffusion}. Let $\tau^{(Z)}:=\inf\{0\leq t \leq T|X_t^{(1)}=\timerev{t}\}$ and:
\begin{equation}\label{eq:SDB_proposal}
 Z_t:=\begin{cases}
X_t^{(1)} & \mbox{ if $0\leq t\leq \tau^{(Z)}$,}\\
\timerev{t} & \mbox{ if $\tau^{(Z)}<t\leq T$.}
\end{cases}
\end{equation}

Then it can be shown that the following equivalence in distribution holds \citep[Theorem 1]{b:bms20}:
\begin{equation}\label{eq:sdb_equiv_in_distro_prop}
Z|\{\tau^{(Z)}\leq T\}\;\overset{d}{=}\;X\bigg|\left\{(X_0=x_0,X_T=x_T),\inf\left\{t\in[0,T]:X_t^{(3)}=X_t\right\}<\infty\right\},
\end{equation}
where $X$ is a solution to \eqref{eq:sde_master} independent of $X^{(3)}$. The law of $Z|\{\tau^{(Z)}\leq T\}$ is denoted by $\mathbb{Z}^{(x_0,x_T,T)}$, and is the law of the \emph{simple diffusion bridge proposal}. Equation \eqref{eq:sdb_equiv_in_distro_prop} shows that $\mathbb{Z}^{(x_0,x_T,T)}$ is equal to the law of the diffusion bridge conditioned on being hit by an independent auxiliary diffusion started at the stationary distribution $\nu$. The methodology for this is summarised in Algorithm \ref{alg:proposal_bridge_sampler}. 
\begin{algorithm}[H]
  \caption{Simple Diffusion Bridge sampler --- Proposal}
  \label{alg:proposal_bridge_sampler}
  \begin{algorithmic}[1]
    \Require{$x_0$, $x_T$, $T$, $b(\cdot)$, $\sigma(\cdot)$}
    \Ensure{Proposal path $Z\sim\mathbb{Z}^{(x_0,x_T,T)}$}
    \While{True}
    \State{Sample $X^{(1)}\sim\mathbb{P}^{(x_0,T)}$}
    \State{Sample $X^{(2)}\sim\mathbb{P}^{(x_T,T)}$}
    \State{Set $\timerev{t}:=X^{(2)}_{T-t}$, $t\in[0,T]$}
    \State{Set $\tau^{(Z)}:=\inf\{0 \leq t\leq T: X^{(1)}_t=\timerev{t}\}$}
    \If{$\tau^{(Z)}<\infty$}
    \State{Set $Z_{t}\leftarrow X_{t}^{(1)}\mathbb{I}_{\{t\leq \tau^{(Z)}\}}+\timerev{t}\mathbb{I}_{\{t>\tau^{(Z)}\}}$, $t\in[0,T]$}
    \State{\Return $Z$}
    \EndIf
    \EndWhile
  \end{algorithmic}
\end{algorithm}

Noting that $\mathbb{Z}^{(x_0,x_T,T)}$ is not the desired law $\mathbb{P}^{(x_0,x_T,T)}$, the second part of the \emph{simple diffusion bridge} sampler accounts for this using a Metropolis--Hastings correction. Letting $\mathcal{C}([0,T])$ denote the canonical space of continuous functions defined on the time interval $[0,T]$, and define $\mathcal{A}_x$ to be the set of functions $y\in\mathcal{C}([0,T])$ that intersect $x\in\mathcal{C}([0,T])$:
\begin{equation*}
\mathcal{A}_x:=\{y\in\mathcal{C}([0,T])|\mbox{gr}(y)\cap \mbox{gr}(x)\neq \emptyset\}.
\end{equation*}
Moreover, let $\mathcal{A}$ represent the set of pairs of intersecting functions: 
\begin{equation*}
\mathcal{A}:=\{(x,y)\in\mathcal{C}([0,T])\times \mathcal{C}([0,T])|y\in\mathcal{A}_x\}.
\end{equation*}
Let $\widetilde{X}$ denote the target diffusion bridge (independent from $X^{(3)}$) and define:
\begin{equation*}
\pi_T(x):=\mathbb{P}(X^{(3)}\in \mathcal{A}_x),\qquad \pi_T:=\mathbb{P}((\widetilde{X},X^{(3)})\in \mathcal{A}).
\end{equation*}
Then as established by \citep[Theorem 2.1]{b:bs14}
\begin{equation*}
\frac{\dd\mathbb{Z}^{(x_0,x_T,T)}}{\dd\mathbb{P}^{(x_0,x_T,T)}}(Z)=\frac{\pi_T(Z)}{\pi_T},
\end{equation*}
which gives an explicit connection between $\mathbb{Z}^{(x_0,x_T,T)}$ and $\mathbb{P}^{(x_0,x_T,T)}$, and provides the correction term in an independence sampler. More precisely, denoting by $Z^{(n)}$ the sample path kept at the $n^{th}$ iteration, we sample a new path $Z\sim\mathbb{Z}^{(x_0,x_T,T)}$ and set $Z^{(n+1)}\leftarrow Z$ with probability:
\begin{equation} \psi(Z^{(n)},Z)=\frac{\frac{\dd\mathbb{Z}}{\dd\mathbb{P}}(Z^{(n)})}{\frac{\dd\mathbb{Z}}{\dd\mathbb{P}}(Z)}\wedge 1=\frac{\pi_T(Z^{(n)})}{\pi_T(Z)}\wedge 1,
\end{equation}
where we suppress $^{(x_0,x_T,T)}$ from the superscripts of the laws for the sake of clarity. Otherwise set $Z^{(n+1)}\leftarrow Z^{(n)}$. Although $1/\pi_T(Z)$ cannot be computed in closed form, it is possible to construct a positive, unbiased estimator. This is sufficient for implementing the algorithm \citep{as:ar09}: by substituting the likelihood for a positive, unbiased estimator of it, the invariant distribution of the Markov Chain is unaltered. For a proposal sample $Z\in \mathcal{C}([0,T])$, one such estimator is
\begin{equation}
  \mathcal{T}:=\min\{i\in\mathbb{N}_+:X^{(3,i)}\in\mathcal{A}_Z\}, \label{AuxT}
\end{equation}
with $X^{(3,i)}$, $i\in\mathbb{N}_+$, i.i.d.\ copies of the auxiliary diffusion $X^{(3)}$. \algoref{alg:metropolis_correction} provides a summary of the full pseudo-marginal independence sampler (including the embedded proposal mechanism detailed in \algoref{alg:proposal_bridge_sampler}). We caution that \algoref{alg:metropolis_correction} is not geometrically ergodic \citep[Theorem 8]{as:ar09}.

\begin{algorithm}[H]
  \caption{Simple Diffusion Bridges --- Full MCMC sampler}
  \label{alg:metropolis_correction}
  \begin{algorithmic}[1]
    \Require{$\{Z^{(n)},\mathcal{T}^{(n)}\}$---output from the previous MCMC step}
    \Ensure{$\{Z^{(n+1)},\mathcal{T}^{(n+1)}\}$---new state of the Markov Chain}
    \State{Sample $Z$ using \algoref{alg:proposal_bridge_sampler} and set $\mathcal{T}\leftarrow 0$}
    \Repeat
    \State{$\mathcal{T} \leftarrow \mathcal{T} + 1$}
    \State{Sample $v\sim\nu$}
    \State{Sample path $X^{(3)}\sim\mathbb{P}^{(T,v)}$}
    \State{Set $\tau^{(aux)}:=\inf\{0\leq t \leq T: X^{(3)}_t=Z_t\}$}
    \Until{$\tau^{(aux)}<\infty$}
    \State{Sample $U\sim \texttt{Unif([0,1])}$}
    \If{$U\leq\frac{\mathcal{T}}{\mathcal{T}^{(n)}}$}
    \Return{$\{Z,\mathcal{T}\}$}
    \Else{ \Return $\{Z^{(n)},\mathcal{T}^{(n)}\}$}
    \EndIf
  \end{algorithmic}
\end{algorithm}

If the simulation of $X^{(1)}$, $\timerev{}$, and $X^{(3)}$ is linear in $T$, then the overall algorithmic complexity of \algoref{alg:metropolis_correction} is also linear in $T$ \cite[Section 2.2]{b:bs14}; \cite{b:bms20}. In practice, and rather surprisingly, the contribution to the cost from the geometric number of trials of the event $\{\tau^{(aux)}<\infty\}$ will typically \emph{decrease} with increasing $T$, as the intersection of $X^{(3)}$ and $Z$ becomes more likely over bridges of longer duration.

For reasons of computational tractability, the implementation of the \emph{simple diffusion bridge} sampler presented by Bladt \& S\o{}rensen \cite{b:bs14} employed a number of approximations. In particular, to simulate $X^{(i)}$, $i=1,2,3$ a discretisation approach based on a stochastic Taylor approximation \citep{bk:kp11} was taken. In addition, a further pragmatic heuristic was used to define the required intersection times of the unconditioned diffusions. The need for similar approximations arises in the independence sampler correction. Consequently these approximations in the simple diffusion bridge sampler propagate, and the desired law $\mathbb{P}^{(x_0,x_T,T)}$ is not recovered exactly. Furthermore, the accuracy of the achieved sampler is difficult to understand. It is these approximations which (for a class of diffusions we introduce fully later) we want to remove from this scheme. Full details are deferred to \secref{sec:CDB}.


\section{Exact and epsilon-strong simulation of unconditioned diffusions}\label{sec:PSRS}

Two sources of error arising in the \emph{simple diffusion bridges} approach outlined in \secref{sec:SDB} are the discretisation errors associated with the simulation of the required diffusion sample paths ($X^{(i)}$, $i\in\{1,2,3\}$), and the determination of the crossing times of the diffusion sample paths. The approach we outline to resolve these errors in \secref{sec:CDB} exploits the availability of methodology to simulate diffusions without time-discretisation error. One class of such methods, which we term \emph{$\varepsilon$-strong simulation}, is appropriate here because it provides auxiliary information about the diffusion including a random collection of compact time-space regions over which the diffusion locally evolves. The flexibility of being able to simulate a trajectory with enfolding local bounds, and to any desired resolution, is useful when determining whether multiple such trajectories intersect, as indicated by the methodology in \secref{sec:SDB} and more fully detailed in \secref{sec:CDB}. For simplicity in this section we will consider a single trajectory, $X$, and denote by $(X^\uparrow(n):n=1,\dots)$ and $(X^\downarrow(n):n=1,\dots)$ sequences of upper and lower bounding processes respectively, which converge to $X$ as the `resolution index' $n\to\infty$. In particular, we require we have almost surely for all $t\in[0,T]$,
\begin{align*}
X^{\downarrow}_t(n) \leq X^{\downarrow}_t(n+1) \leq X_t \leq X^{\uparrow}_t(n+1) \leq X^{\uparrow}_t(n),
\end{align*}
with 
\begin{align}
\lim_{n\to\infty}\,\sup_{t\in[0,T]} \Big|X^{\uparrow}_t(n) - X^{\downarrow}_t(n)\Big| \to 0, \qquad a.s. \label{eq:epsilonstrong}
\end{align}

It is possible to do this by means of Pollock et al.~\cite[Sections 3.2 \& 5]{b:pjr16}, which we will outline in the remainder of this section. We first describe how to simulate \emph{exactly} (in the sense that it is without time-discretisation error) the trajectory of a class of diffusions at any finite collection of time points, by means of rejection and using a Brownian motion proposal. We then consider how to find appropriate upper and lower bounding processes of the type described in \eqref{eq:epsilonstrong}. 

For Brownian motion to be an appropriate proposal, we first consider a transformation of \eqref{eq:sde_master} to one with unit volatility; this is always possible for a scalar diffusion using the \emph{Lamperti} transformation \citep[\S IV.4]{bk:kp11}. In particular, if
\begin{equation}\label{eq:lamperti}
\eta(x):=\int^{x}\frac{1}{\sigma(u)}du,
\end{equation}
then $Y_t:=\eta(X_t)$ solves the following SDE:
\begin{equation}\label{eq:diff_unit_vola}
\dd Y_t = \alpha(Y_t)\dd t + \dd W_t, \qquad Y_0 = \eta(x_0), \qquad t\in[0,T],
\end{equation}
where
\begin{equation*}
\alpha(y) := \frac{b}{\sigma}\left(\eta^{-1}(y)\right) - \frac{1}{2} \sigma'\left(\eta^{-1}(y)\right).
\end{equation*}
As such we will assume without loss of generality that the diffusion coefficient $\sigma$ in \eqref{eq:sde_master} is identically equal to $1$. We impose a number of other standard regularity conditions on $\alpha$ as given in \citep[page 1080]{b:bpr06}, which includes that $\alpha$ is continuously differentiable.

Under these conditions, by application of \citep[Algorithm 4]{b:pjr16} it is possible to simulate finite dimensional representations of diffusion sample paths, such as those solving \eqref{eq:diff_unit_vola}, \emph{exactly} (without discretisation error). In particular, the output of the algorithm is a \emph{skeleton},
\begin{align} \label{eq:ea_skeleton}
    \mathcal{S}(X) = \left\{\left(\chi_i,X_{\chi_i}\right)^{\kappa+1}_{i=0}, \left(R_X^{[\chi_{i},\chi_{i+1}]}\right)^{\kappa}_{i=0}\right\},
\end{align}
where $\kappa$ is a Poisson distributed random variable determining the number of uniformly distributed time points, $\chi_1,\dots,\chi_\kappa$, between the end-points ($\chi_0:=0$ and $\chi_{\kappa+1}:=T$), at which the trajectory is revealed.  $R_X^{[\chi_{i},\chi_{i+1}]} := \{[L^{\downarrow}_{X[\chi_j,\chi_{j+1}]},L^{\uparrow}_{X[\chi_j,\chi_{j+1}]}], [U^{\downarrow}_{X[\chi_j,\chi_{j+1}]},U^{\uparrow}_{X[\chi_j,\chi_{j+1}]}]\}$ is a function of the sample path $X$, over the time interval $[\chi_i,\chi_{i+1}]$, providing a pair of intervals in which the minimum and maximum of the trajectory are constrained ($[L^{\downarrow}_{X[\chi_j,\chi_{j+1}]},L^{\uparrow}_{X[\chi_j,\chi_{j+1}]}]$ and $[U^{\downarrow}_{X[\chi_j,\chi_{j+1}]},U^{\uparrow}_{X[\chi_j,\chi_{j+1}]}]$ respectively). For the purposes of the methodology developed in this paper we require only $L^{\downarrow}_{X[\chi_j,\chi_{j+1}]}$ and $U^{\uparrow}_{X[\chi_j,\chi_{j+1}]}$ (which determines a compact interval which constrains the trajectory).

\begin{figure}[!ht]
\begin{center}
\subfigure[Initialisation.
\label{subfig:conf1p1}]{\includegraphics[width=0.45\textwidth]{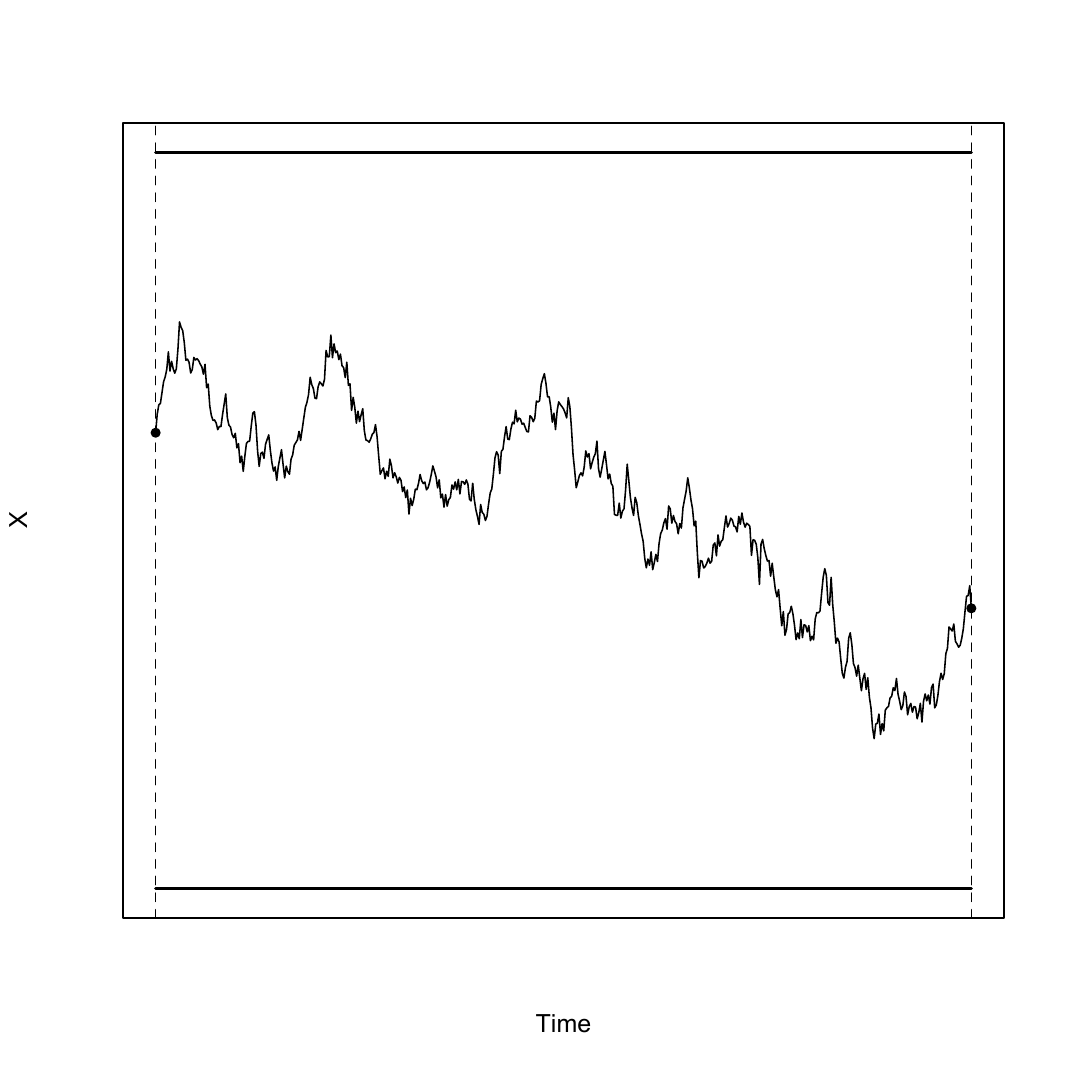}}
\hfill
\subfigure[Iteration 2. 
\label{subfig:conf1p2}]{\includegraphics[width=0.45\textwidth]{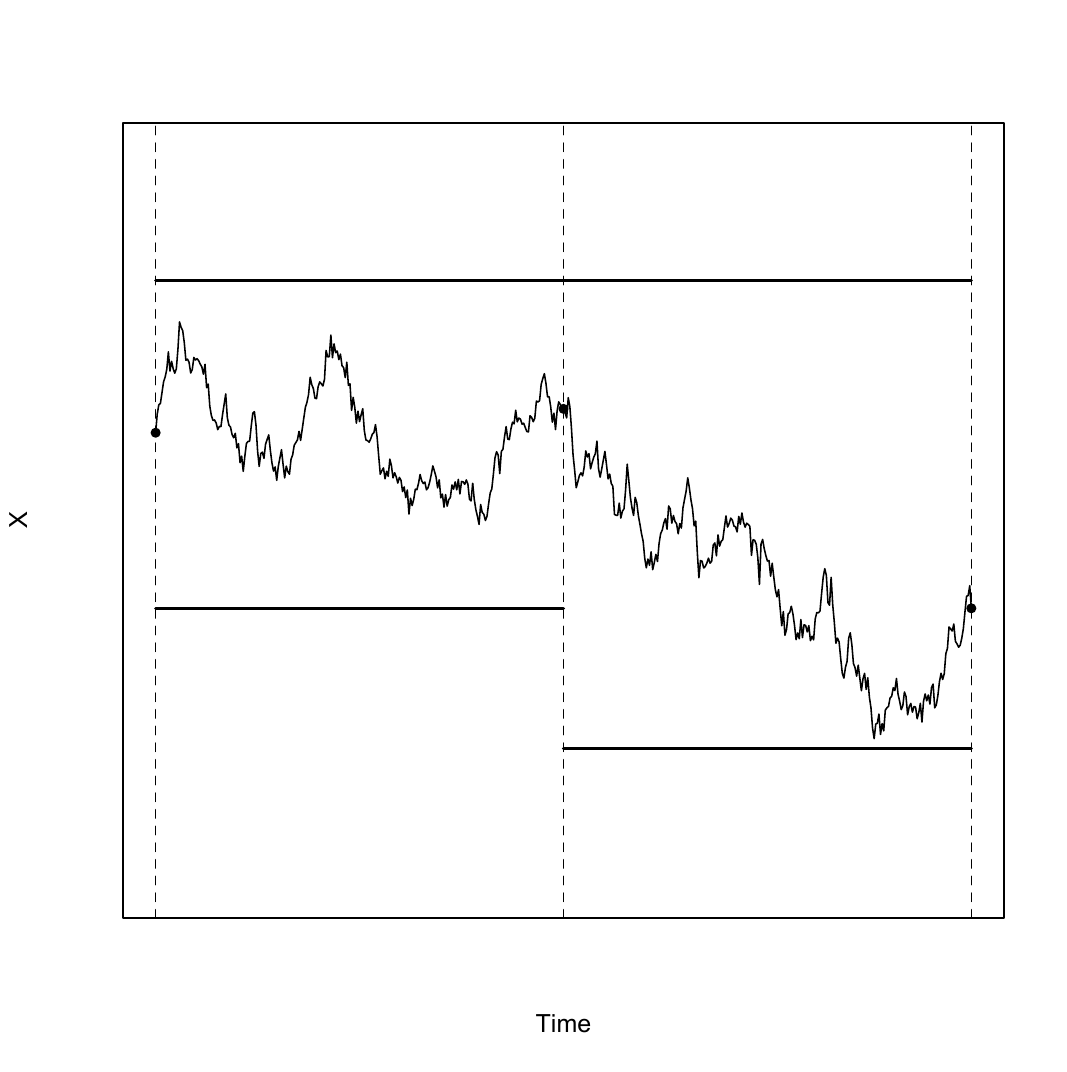}}
\hfill
\subfigure[Iteration 3.  
\label{subfig:conf1p3}]{\includegraphics[width=0.45\textwidth]{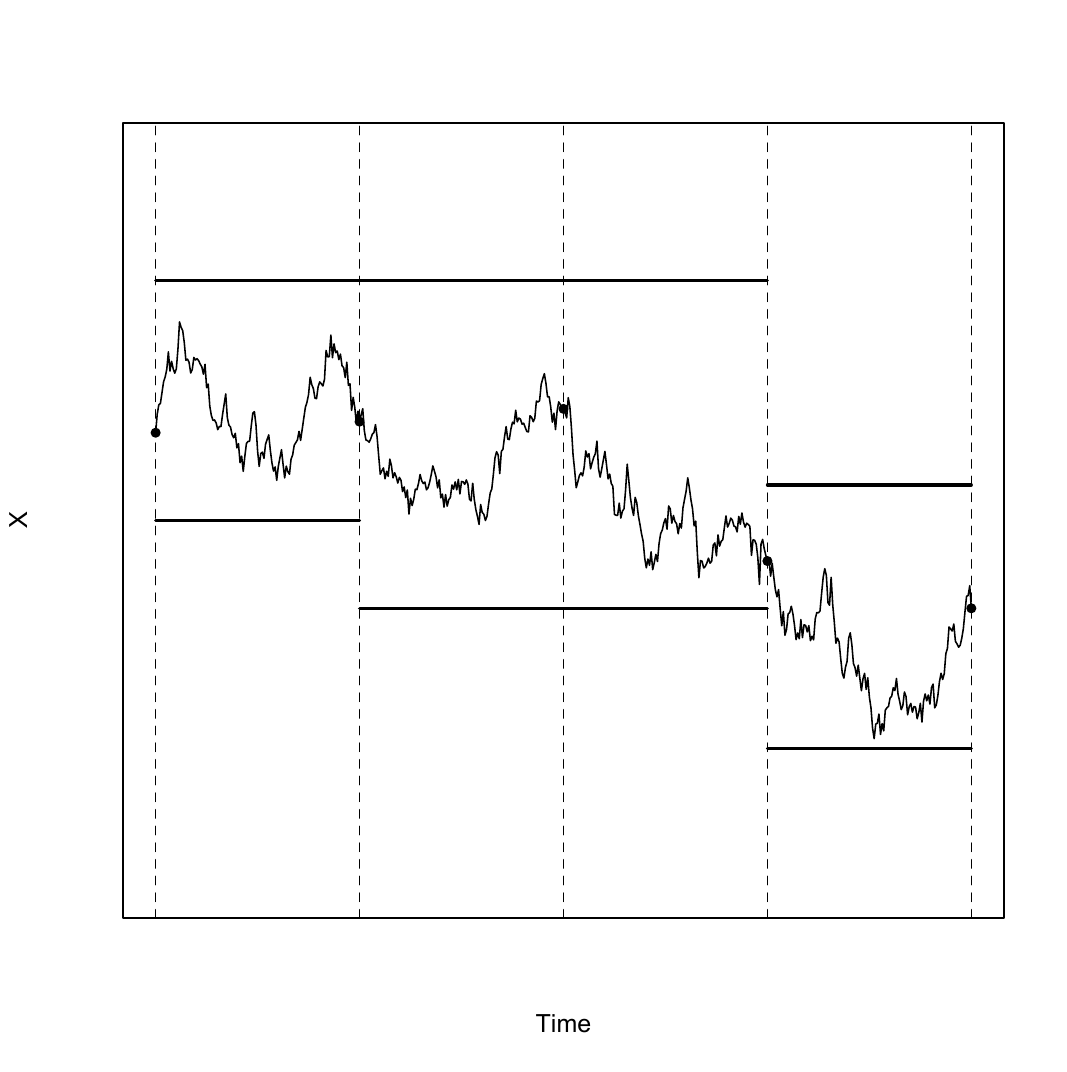}}
\hfill
\subfigure[Iteration 4. 
\label{subfig:conf1p4}]{\includegraphics[width=0.45\textwidth]{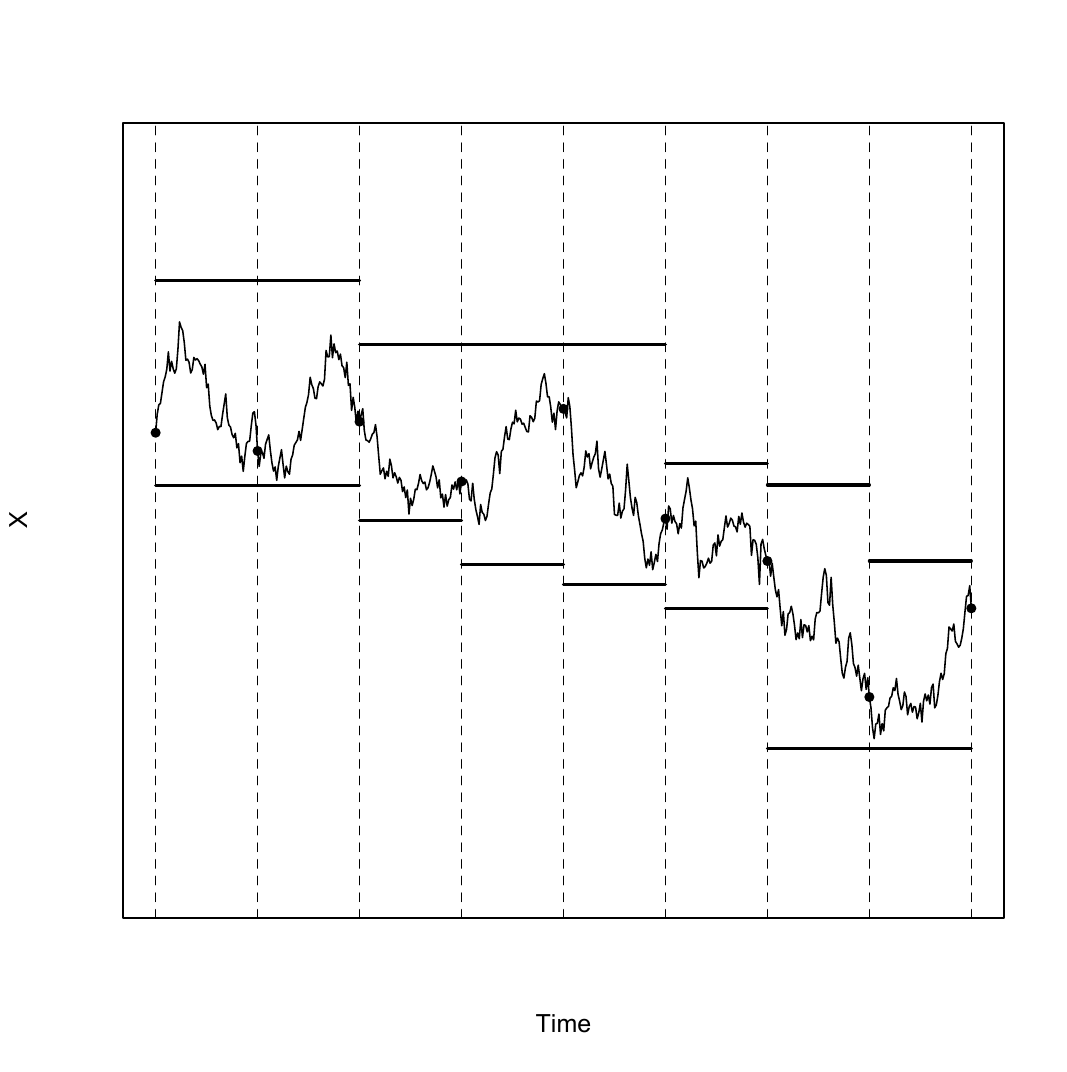}}
\caption{An illustrative diffusion trajectory in an interval $[\chi_j,\chi_{j+1}]$, together with its skeleton at initialisation (\figref{subfig:conf1p1}) and at increasing resolutions (Figures \ref{subfig:conf1p2} -- \ref{subfig:conf1p4}). The skeleton at a given resolution is given by the trajectory's temporal location (black dots) at a collection of times (black dashed vertical lines) and almost sure bounds between consecutive times (black solid horizontal lines).} \label{fig:some_epsilon_strong}
\end{center}
\end{figure}

By means of \citep[Algorithm 22]{b:pjr16} it is possible to retrospectively reveal the realised sample path at any additional time point, conditional on the skeleton \eqref{eq:ea_skeleton}, noting that the unrealised trajectory has the law of a Brownian bridge between each pair of skeleton points conditioned on the associated layer. It is then possible to augment the skeleton with additional layer information to account for the fact that an existing layer has been bisected by the new time point. In the later sections of this paper we will consider revealing the path, and augmenting the skeleton, at multiple additional time points. This can be done in an iterative manner, at a resolution sufficient for the user (for instance, some $\epsilon$-tolerance). An illustrative example of a realised skeleton of a trajectory following from the entire $\epsilon$-strong simulation approach can be found in \figref{fig:some_epsilon_strong}. 


\section{Confluent diffusion bridges}\label{sec:CDB}

In this section we outline our \emph{confluent diffusion bridge} approach, which incorporates a discretisation free path-space rejection sampler for an unconditioned ergodic diffusion using $\epsilon$-strong simulation as described in \secref{sec:PSRS}, within the \emph{simple diffusion bridge} framework of Bladt \& S\o{}rensen as outlined in \secref{sec:SDB}. Here we carry over the assumptions from both of these areas: in particular, from \secref{sec:SDB} we assume that the unconditioned diffusion is ergodic with unique stationary distribution $\nu$, and from \secref{sec:PSRS} we assume the regularity conditions on $\alpha$ as described in \citep[page 1080]{b:bpr06}.


\subsection{Sampling proposals} \label{sec:CDB_prop}

\begin{figure}[!ht]
\begin{center}
\subfigure[Initial skeletons for $X^{(1)}$ and $\timerev{}$.
\label{subfig:conf2p1}]{\includegraphics[width=0.45\textwidth]{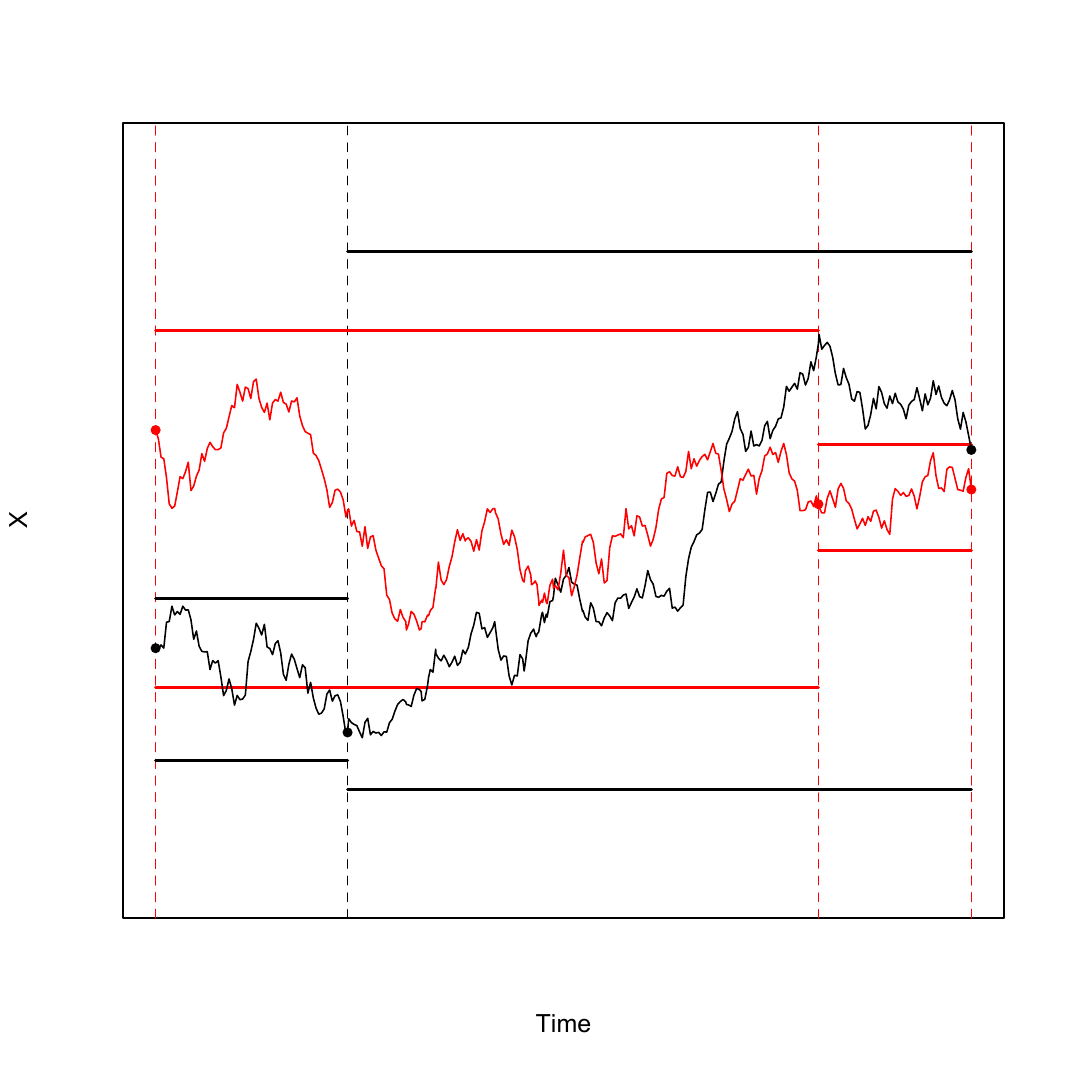}}
\hfill
\subfigure[Cross-population of $X^{(1)}$ and $\timerev{}$ so they share a common temporal resolution (given by blue dashed vertical lines). 
\label{subfig:conf2p2}]{\includegraphics[width=0.45\textwidth]{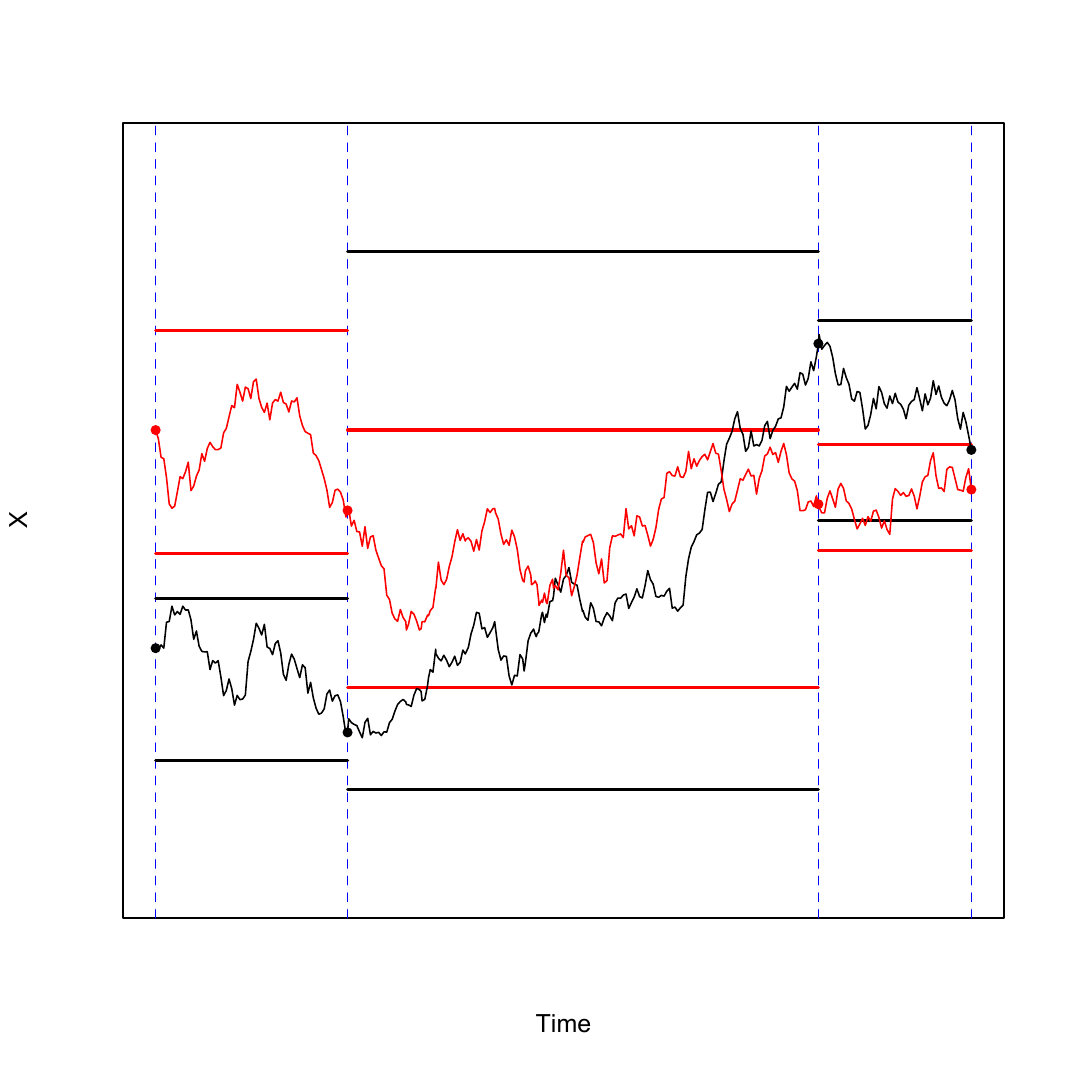}}
\hfill
\subfigure[Selective resolution of the skeletons of $X^{(1)}$ and $\timerev{}$ to determine if they cross, and if so the interval in which they do cross (given by blue dashed vertical lines). 
\label{subfig:conf2p3}]{\includegraphics[width=0.45\textwidth]{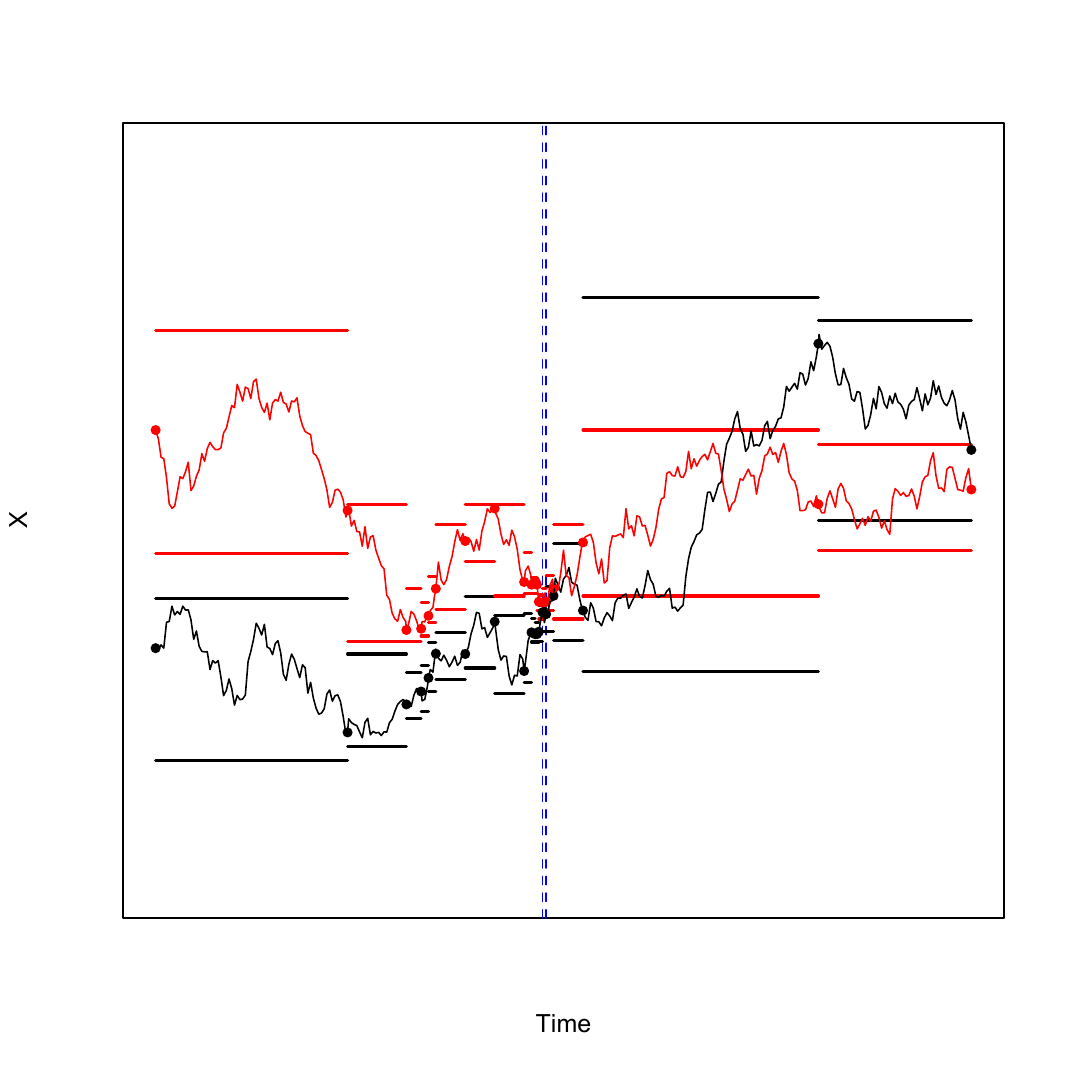}}
\hfill
\subfigure[Construction of the skeleton of $Z$ by concatenating the skeletons of $X^{(1)}$ and $\timerev{}$ around the (first) intersection interval.
\label{subfig:conf2p4}]{\includegraphics[width=0.45\textwidth]{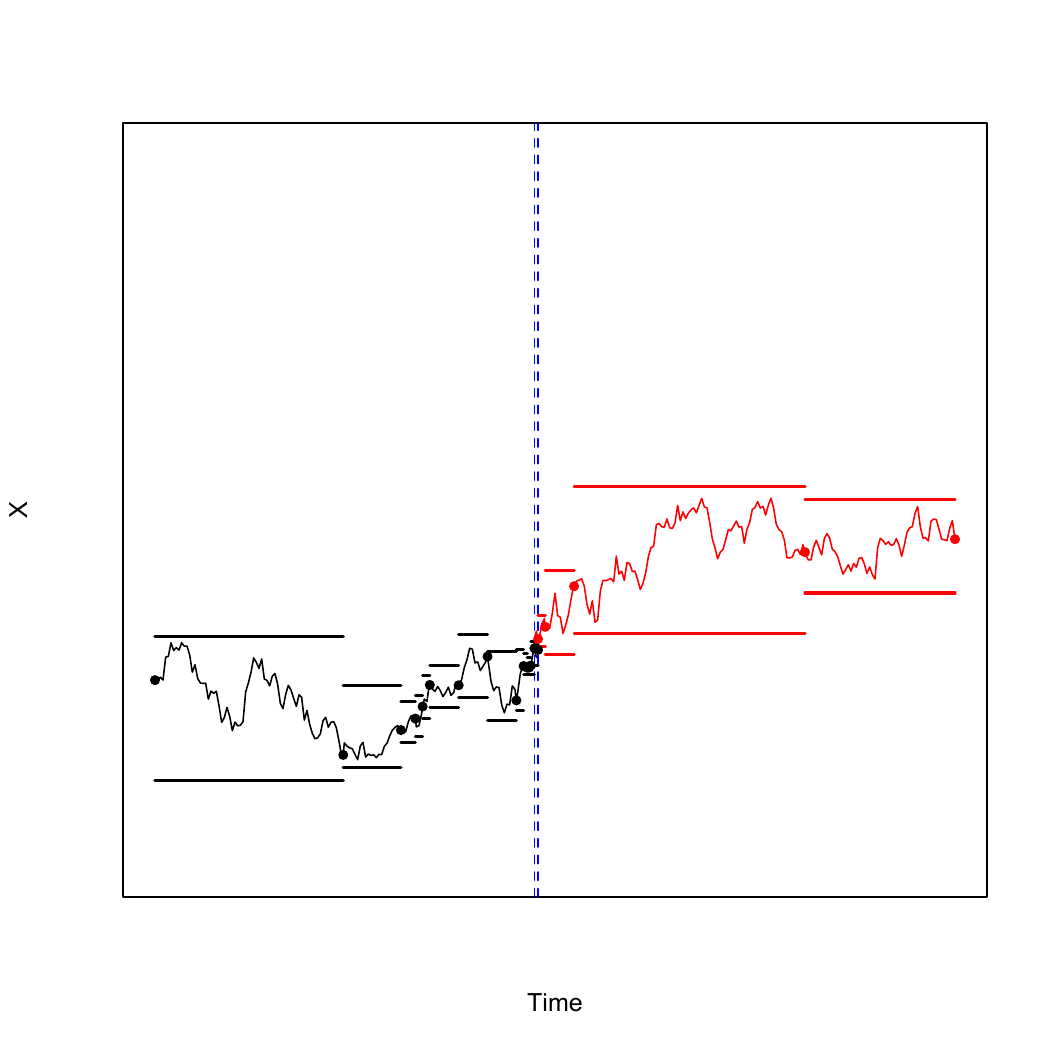}}
\caption{Illustrative diffusion trajectories given by $X^{(1)}$ (black), $\timerev{}$ (red), and the resulting proposal $Z$.} \label{fig:prop_sim}
\end{center}
\end{figure}

Our aim is to sample from $\mathbb{Z}^{(x_0,x_T,T)}$, avoiding the discretisation errors in the \emph{simple diffusion bridge} approach, by using the exact and  $\varepsilon$-strong simulation approach outlined in \secref{sec:PSRS}. Recalling the construction of the proposal diffusion bridge $Z$ as given in \eqref{eq:SDB_proposal} in the \emph{simple diffusion bridge} approach, it is itself composed of two unconditioned diffusions: A \emph{forward} diffusion, $X^{(1)}$, initialised at the left hand end-point $x_0$, to some time $\tau^{(Z)}$; and a \emph{time-reversed diffusion}, $\timerev{}$, from $\tau^{(Z)}$ to the right hand end-point $x_T$. Here  $\tau^{(Z)}=\inf\{0\leq t \leq T|X_t^{(1)}=\timerev{t}\}$ is simply the first intersection time of $X^{(1)}$ and $\timerev{}$. A rejection sampling scheme to sample from $\mathbb{Z}^{(x_0,x_T,T)}$ would simply sample pairs of candidate paths $X^{(1)}$ and $\timerev{}$ until a pair is obtained that intersects (i.e.\ $\tau^{(Z)}<\infty$). \figref{fig:prop_sim} illustrates the procedure graphically.


We begin by simulating skeletons for $X^{(1)}$ and $X^{(2)}$ (denoted by $\mathcal{S}(X^{(1)})$ and $\mathcal{S}(X^{(2)})$ respectively), which immediately provides the skeleton $\mathcal{S}(\timerev{})$ for $\timerev{t} = X^{(2)}_{T-t}$ (as shown in \figref{subfig:conf2p1}). These skeletons are then \emph{`cross-populated'}, by revealing the trajectories (including layer information) at additional time points, as described in \secref{sec:PSRS}. In particular, $\mathcal{S}(X^{(1)})$ is cross-populated by the temporal points $\{T-\chi^{(2)}_j\}_{j=1}^{\kappa^{(2)}}$, and $\mathcal{S}(\timerev{})$ is cross-populated by the temporal points $\{\chi^{(1)}_j\}_{j=1}^{\kappa^{(1)}}$. Each resulting skeleton is thus revealed on the common grid
\[
\{\chi_j\}_{j=0}^{\kappa+1}:=\{\chi^{(1)}_j\}_{j=0}^{\kappa^{(1)}+1}\cup \{T-\chi^{(2)}_j\}_{j=0}^{\kappa^{(2)}+1},
\]
(where $\kappa:= \kappa^{(1)}+\kappa^{(2)}$) as summarised by \figref{subfig:conf2p2}.

Next, we must determine $\tau^{(Z)}$ for the two diffusions $X^{(1)}$ and $\timerev{}$, in order to determine whether the candidate for $\mathbb{Z}^{(x_0,x_T,T)}$ is to be accepted. Cross-populating the skeletons as described above so they share a common time grid, which for simplicity we assume to be sorted in increasing order, we can simply sequentially consider the intervals $[\chi_j, \chi_{j+1}]$, $j=0,\dots,\kappa$, querying each interval with the aid of Table \ref{tab:sequential_eval}. Throughout the remainder of this section we will assume, without loss of generality, that $X^{(1)}_0 < \timerev{0}$, and we denote $L^{(2)}_{\chi_j}:=L^{\downarrow}_{\timerev{}[\chi_j,\chi_{j+1}]}$ and $U^{(1)}_{\chi_j}:=U^{\downarrow}_{X^{(1)}[\chi_j,\chi_{j+1}]}$.
\begin{table}
\begin{tabular}{cc|p{0.35\textwidth}}
& Scenario & Interpretation \\\hline
(i) & $U^{(1)}_{\chi_j} < L^{(2)}_{\chi_j}$  & The two paths cannot cross\\[5pt]
(ii) & $X^{(1)}_{\chi_{j+1}} > \timerev{\chi_{j+1}} $ & The two paths must cross\\[5pt]
(iii) & Otherwise & Undetermined
\end{tabular}
\caption{Possible cases in sequential querying of $\mathcal{S}(X^{(1)})$ and $\mathcal{S}(\timerev{})$ to determine whether crossing has occurred, when dealing with the case $X^{(1)}_0 < \timerev{0}$. If $X^{(1)}_0 > \timerev{0}$ then crossing can be determined in the same manner by swapping the superscripts. \label{tab:sequential_eval}}
\end{table}

Considering the current interval (say, $[\chi_j, \chi_{j+1}]$) in our sequential querying, and referring to Table \ref{tab:sequential_eval}, if it falls into case (i) then we simply proceed to the next interval ($[\chi_{j+1}, \chi_{j+2}]$)). In case (ii) we halt our sequential evaluation, having identified that the current interval is one in which the first crossing has occurred (so $\tau^{(Z)}\in[\chi_j, \chi_{j+1}]$). In case (iii) further information is required to determined whether crossing has occurred. We propose to exploit the fact that it is possible to retrospectively reveal additional points from the  paths underlying the existing skeletons (as described in \secref{sec:PSRS}). In particular, we begin by revealing both paths at the mid-point $(\chi_j+\chi_{j+1})/2$ of the current interval, and then augmenting both skeletons with corresponding layer information for interval $[\chi_j, (\chi_j+\chi_{j+1})/2]$ and $[(\chi_j+\chi_{j+1})/2, \chi_{j+1}]$. We then resume our sequential evaluation (continuing with an abuse of notation to write $\{\chi_j\}_{j=0}^{\kappa+1}$ for the set of time points including any additionally revealed ones), starting at the interval $[\chi_j, (\chi_j+\chi_{j+1})/2]$. If upon evaluating all intervals in the sequential manner indicated we find that all intervals fall into case (i), then we can determine that no crossing of $X^{(1)}$ and $\timerev{}$ occurs ($\tau^{(Z)}=\infty$). 

Having simulated a skeleton for $X^{(1)}$ and $\timerev{}$ and determined whether they cross, we can determine whether the candidate for $\mathbb{Z}^{(x_0,x_T,T)}$ is to be accepted. If they do not cross, then we repeat the procedure above until a pair $X^{(1)}$ and $\timerev{}$ are obtained that do cross. If they do cross (as illustrated in \figref{subfig:conf2p3}) then we can simply return a skeleton for $Z$ as suggested by \eqref{eq:SDB_proposal} (as shown in \figref{subfig:conf2p4}). $\mathcal{S}(Z)$ is a more complicated object, and some care is required in its construction as we have only identified the \emph{interval} in which the crossing has occurred and not $\tau^{(Z)}$ itself. As a consequence, we need to retain both candidate skeletons for the interval in which crossing has occurred in order to conditionally reveal $Z$ if required in this interval (as we describe later). For clarity, we provide an expression for $\mathcal{S}(Z)$ which, following the procedure evaluated above, will be resolved more finely than either of the original skeletons $\mathcal{S}(X^{(1)})$ or $\mathcal{S}(\timerev{})$. For simplicity, we denote $\{0=:\xi_{(0)},\xi_{(1)},\dots,\xi_{(m)},\xi_{(m+1)}:=T\}$ as the new, finer, ordered common temporal grid associated with each skeleton following the cross-population procedure used in determining crossing. We further denote $m^{(1)}:=|\{\xi_{(0)},\dots\}\cap[0,\tau^{(Z)}]|$. 
Note from our notation that $\tau^{(Z)}\in[\xi_{(m^{(1)})},\xi_{(m^{(1)}+1)}]$, and we retain within $\mathcal{S}(Z)$ the layer information for both $X^{(1)}$ and $\timerev{}$ corresponding to that interval. Then 
\begin{align}
\mathcal{S}(Z) 
& := \Bigg\{\left(\xi_i,X^{(1)}_{\xi_i}\right)^{m^{(1)}+1}_{i=0}, \left(R_{X^{(1)}}^{[\xi_{i},\xi_{i+1}]}\right)^{m^{(1)}}_{i=0}\Bigg\} \nonumber \\
& \qquad \bigcup \Bigg\{\left(\xi_i,\timerev{\xi_i}\right)^{m+1}_{i=m^{(1)}}, \left(R_{\timerev{}}^{[\xi_{i},\xi_{i+1}]}\right)^{m}_{i=m^{(1)}}\Bigg\}, \label{eq:Z_skeleton}
\end{align}

We summarise the construction of the skeleton of the proposal within our CDB methodology, which we have described above, in \algoref{alg:sampling_Z}.

\begin{algorithm}[H]
  \caption{Confluent Diffusion Bridge --- Proposal \label{alg:sampling_Z}}
  \begin{algorithmic}[1]
    \Require{$x_0$, $x_T$.}
    \Ensure{$\mathcal{S}(Z)$.}
    \While{True}
        \State{Sample $\mathcal{S}(X^{(1)})$ and $\mathcal{S}(\timerev{})$ as per \secref{sec:PSRS}.}
        \State{Cross-populate $\mathcal{S}(X^{(1)})$ and $\mathcal{S}(\timerev{})$ so they share common temporal resolution $\{\chi_j\}_{j=0}^{\kappa+1}$ as per \citep[Algorithm 22]{b:pjr16}.}
        \State{Set $j\leftarrow 0$, $m\leftarrow\kappa$, and $\{\xi_j\}^{m+1}_{j=0}\leftarrow \{\chi_j\}^{m+1}_{j=0}$.}
        \While{$\xi_{(j)}<T$}
            \If{$\Big(X^{(1)}_0 < \timerev{0} \text{ and } U_{\xi_{(j)}}^{(1)} < L_{\xi_{(j)}}^{(2)}\Big)$ or $\Big(X^{(1)}_0 > \timerev{0} \text{ and } U_{\xi_{(j)}}^{(2)} < L_{\xi_{(j)}}^{(1)}\Big)$}
                \item[] \Comment{Table \ref{tab:sequential_eval} Case (i)}
                \State{$j\leftarrow j+1$.}
        \ElsIf{$\Big(X^{(1)}_0 < \timerev{0} \text{ and } X^{(1)}_{\xi_{(j+1)}} > \timerev{\xi_{(j+1)}}\Big)$ or $\Big(X^{(1)}_0 > \timerev{0} \text{ and } X^{(1)}_{\xi_{(j+1)}} < \timerev{\xi_{(j+1)}}\Big)$}
                \item[] \Comment{Table \ref{tab:sequential_eval} Case (ii)}
                \State{Set $m^{(1)}\leftarrow j$.} 
                \State{\Return $\mathcal{S}(Z)$ as per \eqref{eq:Z_skeleton}.}        
            \Else 
                \item[] \Comment{Table \ref{tab:sequential_eval} Case (iii)}
                \State{Set $\xi^*\leftarrow (\xi_{(j)}+\xi_{(j+1)})/2$, $m\leftarrow m+1$, and $\{\xi_j\}^{m+1}_{j=0}\leftarrow \{\{\xi_j\}\cup\xi^*\}$.}
                \State{Update $\mathcal{S}(X^{(1)})$ and $\mathcal{S}(\timerev{})$ to include $\xi^*$ as per \citep[Algorithm 22]{b:pjr16}.}        
            \EndIf
        \EndWhile
    \EndWhile
  \end{algorithmic}
\end{algorithm}


\subsection{Embedding the proposal within Markov chain Monte Carlo} \label{sec:CDB_ar}

\begin{figure}[ht]
\begin{center}
\subfigure[Initial skeleton of (non-crossing) auxiliary path $X^{(3)}$.
\label{subfig:conf3p1}]{\includegraphics[width=0.4\textwidth]{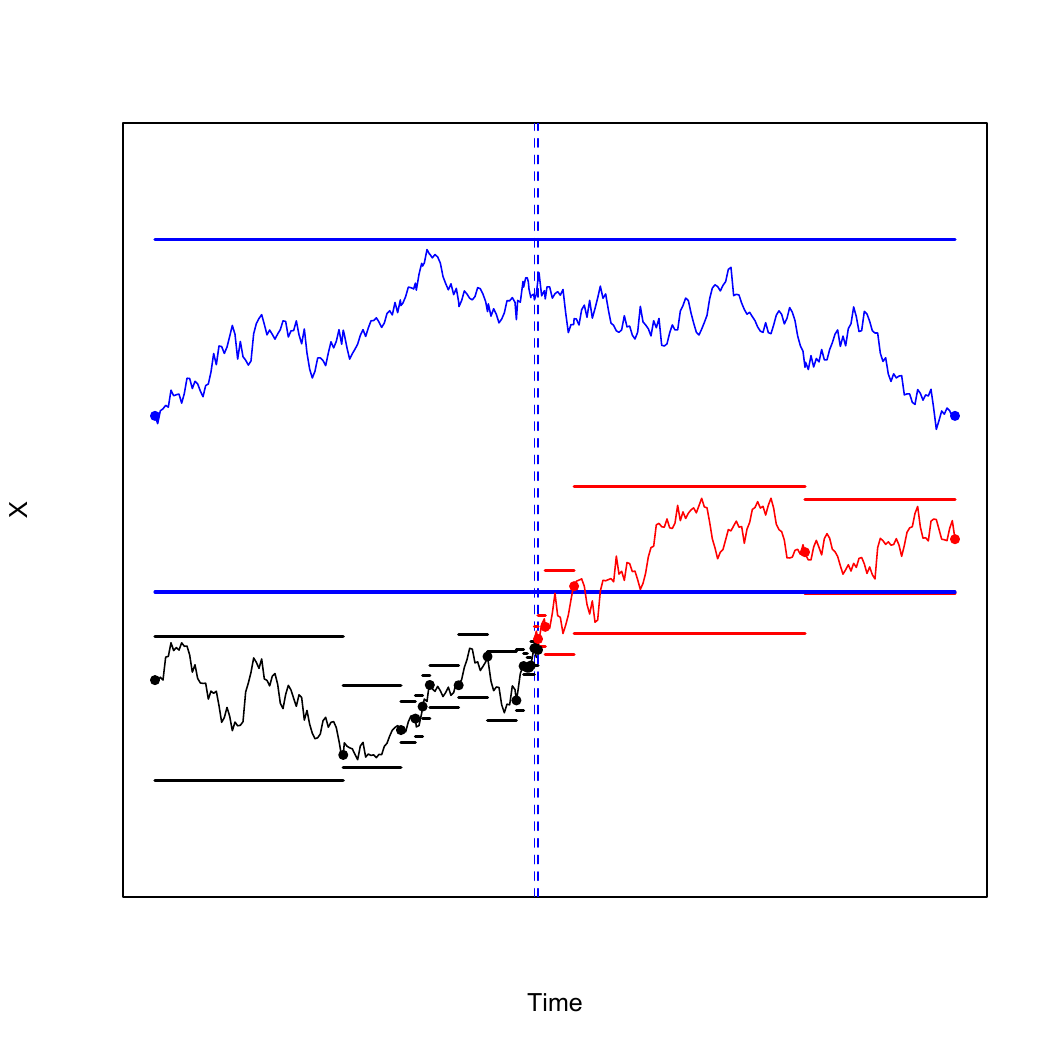}}
\hfill
\subfigure[Skeleton of (non-crossing) auxiliary path $X^{(3)}$, at a resolution sufficient to show no crossing occurs.
\label{subfig:conf3p2}]{\includegraphics[width=0.4\textwidth]{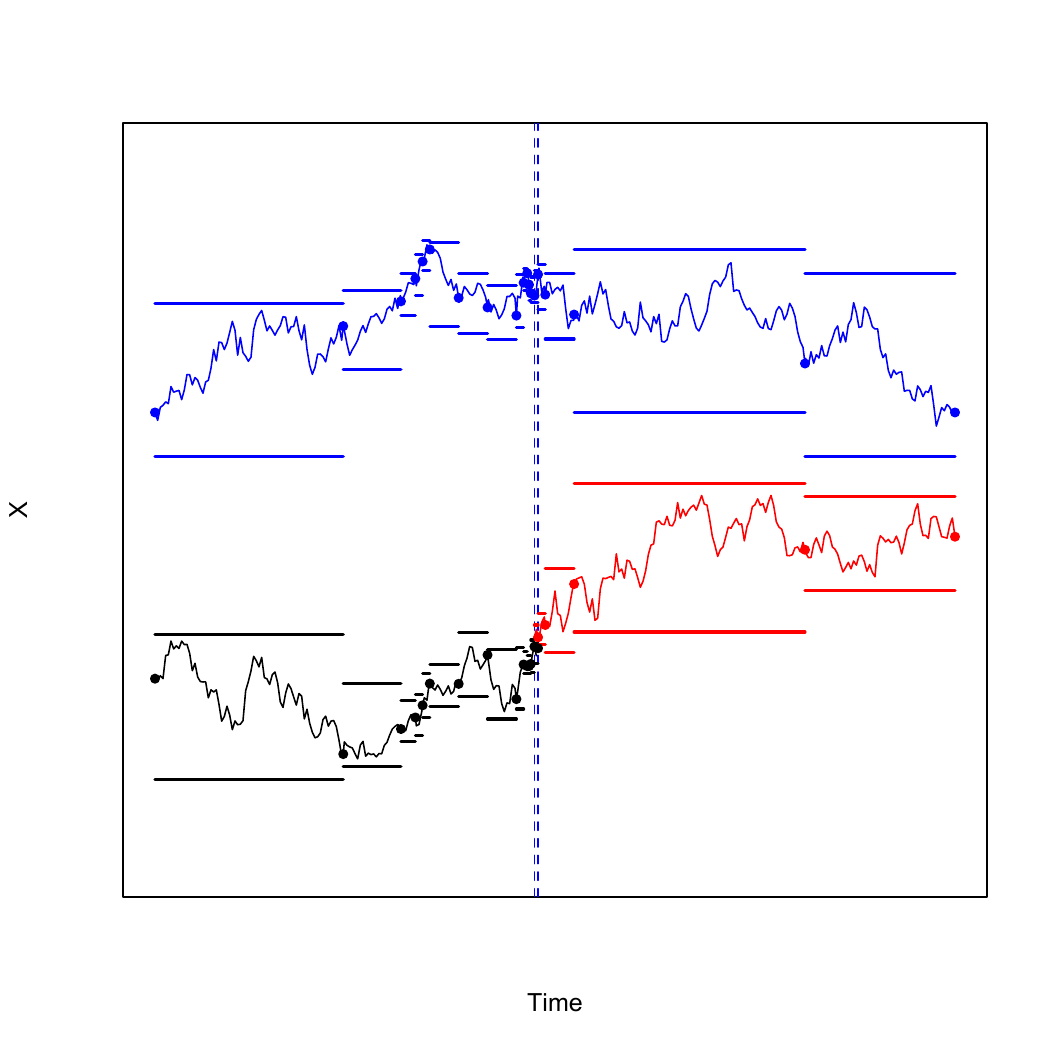}}
\hfill
\subfigure[Initial skeleton of (crossing) auxiliary path $X^{(3)}$.
\label{subfig:conf3p3}]{\includegraphics[width=0.4\textwidth]{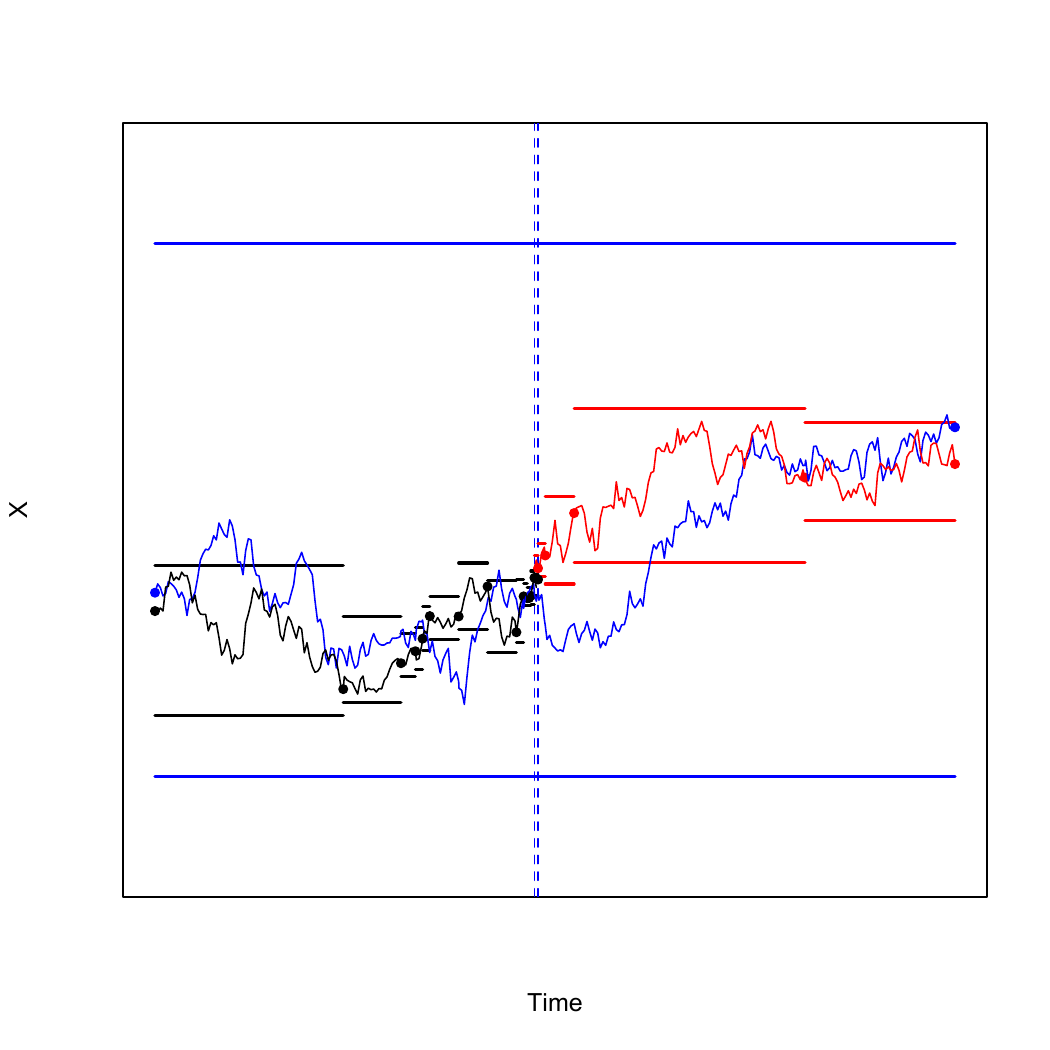}}
\hfill
\subfigure[Skeleton of (crossing) auxiliary path $X^{(3)}$, at a resolution which shows crossing occurs at an interval which does not coincide with the intersection interval of ${\cal S}(Z)$. 
\label{subfig:conf3p4}]{\includegraphics[width=0.4\textwidth]{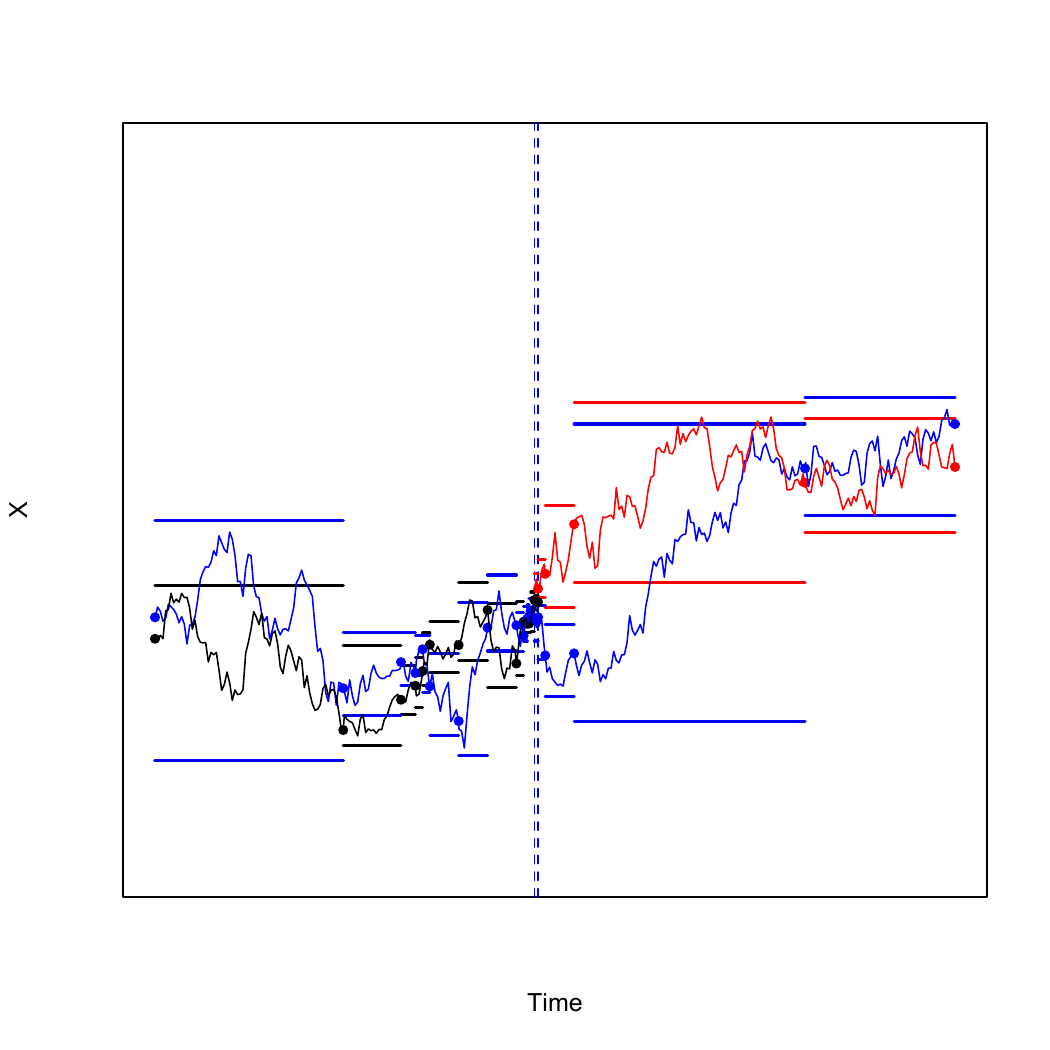}}
\caption{Illustrative proposal trajectory $Z$ from \figref{fig:prop_sim}, together with illustrative trajectories of auxiliary paths $X^{(3)}$ (blue) in two scenarios: Figures \ref{subfig:conf3p1} and \ref{subfig:conf3p2} consider a trajectory $X^{(3)}$ when crossing does not occur; Figures \ref{subfig:conf3p3} and \ref{subfig:conf3p4} consider the case where crossing does occur. The vertical blue dashed lines again indicate the (first) intersection interval of ${\cal S}(X^{(1)})$ and ${\cal S}(\timerev{})$ which comprise ${\cal S}(Z)$ and which requires careful consideration.} \label{fig:refine_prop}
\end{center}
\end{figure}

To determine whether the proposal skeleton from \algoref{alg:sampling_Z}, $\mathcal{S}(Z)$, is accepted or rejected, then following the simple diffusion bridge approach outlined in \secref{sec:SDB} we repeatedly simulate skeletons of auxiliary diffusion paths $X^{(3)}$ until we can determine that one intersects with $Z$. This is illustrated in \figref{fig:refine_prop}. Recall that $\mathcal{T}$ denotes the number such paths needed, as given by \eqref{AuxT}.

For each realisation of $X^{(3)}$ we again exploit the $\varepsilon$-strong simulation we introduced in \secref{sec:PSRS}. To do so we simply initialise the trajectory by simulating from $\nu$, and then apply \citep[Algorithm 4]{b:pjr16} as in \secref{sec:PSRS} to obtain $\mathcal{S}(X^{(3)})$. To determine whether the auxiliary path has crossed the proposal $Z$, we begin by cross-populating $\mathcal{S}(Z)$ and $\mathcal{S}(X^{(3)})$ so that they share a common set of temporal points (and associated layers) by application of \citep[Algorithm 22]{b:pjr16}. 

Following this cross-population it is important we do not lose track of the interval containing $\tau^{(Z)}$. If a new temporal point has been added in the time interval containing $\tau^{(Z)}$, we run Algorithm \ref{alg:sampling_Z} in this interval to identify the subinterval to which $\tau^{(Z)}$ belongs. Subsequently we populate $\mathcal{S}(X^{(3)})$ with any additional temporal points in order to maintain a common resolution.

To determine whether the auxiliary path $X^{(3)}$ crosses the proposal path $Z$ or not, that is whether for
\[
\tau^{(aux)}:=\inf\{0\leq t \leq T: X^{(3)}_t=Z_t\}
\]
we have $\tau^{(aux)}<\infty$ or $\tau^{(aux)}=\infty$, we can now replicate the procedure outlined in \secref{sec:CDB_prop}. We proceed in \emph{almost} the same fashion as we did for determining whether $X^{(1)}$ and $\timerev{}$ crossed: As presented in \secref{sec:PSRS}, $\varepsilon$-strong simulation can be used to determine whether two diffusion paths cross one another over a given interval. The only difference is should we need to determine whether the auxiliary path crosses the proposal path in the interval containing $\tau^{(Z)}$, then to apply \secref{sec:PSRS} we first temporally bisect (iteratively if necessary) both $\mathcal{S}(Z)$ and $\mathcal{S}(X^{(3)})$.

Having developed a procedure to determine whether $X^{(3)}$ and $Z$ cross, we can now return to the \emph{simple diffusion bridge} approach of Bladt \& S\o{}rensen \cite{b:bs14}, outlined in \secref{sec:SDB}: a proposal $Z$ is employed within a pseudo-marginal independence sampler for which $1/\pi_T(Z)$ appearing in the Metropolis--Hastings ratio is estimated by $\mathcal{T}$ in \eqref{AuxT}. We summarise this in \algoref{alg:cdb_alg}. For presentation we resolve whether crossing has occurred in temporal order, but this is not necessary---any crossing will do---and therefore the search can often be accelerated by looking for ``trivial'' crossings first, before having to employ any refinements of layers.

Note that in determining whether the auxiliary path crossed the proposal, we have revealed the proposal path at additional time-points. Importantly, we never discard any additional information that has been sampled for the proposal path $Z$. If after sampling the auxiliary path it turns out that the two trajectories did not cross, we must keep all information about the path $Z$ and carry it into the next experiment in which we sample the subsequent auxiliary path. It is the $\varepsilon$-strong simulation approach we employ that allows us to retain all pertinent information of the previously simulated auxiliary paths without a corresponding increase in computational cost at each iteration.

\begin{algorithm}[H]
  \caption{Confluent Diffusion Bridge --- Full MCMC sampler}
  \label{alg:cdb_alg}
  \begin{algorithmic}[1]
    \Require{$\{\mathcal{S}(Z)^{(n)},\mathcal{T}^{(n)}\}$---output from the previous MCMC step}
    \Ensure{$\{\mathcal{S}(Z)^{(n+1)},\mathcal{T}^{(n+1)}\}$---new state of the Markov Chain}
    \State{Sample $\mathcal{S}(Z)$ given in \eqref{eq:Z_skeleton} as per \algoref{alg:sampling_Z}.}
    \State{Set $\mathcal{T}\leftarrow 0$ and $\widetilde{m}\leftarrow m$.}
    \Repeat
    \State{$\mathcal{T} \leftarrow \mathcal{T} + 1$}
    \State{Sample path $\mathcal{S}(X^{(3)})$ as per \secref{sec:PSRS}, initialised at $X^{(3)}_0\sim\nu$.}
    \State{Cross-populate $\mathcal{S}(Z)$ and $\mathcal{S}(X^{(3)})$ so they share common temporal resolution as per \citep[Algorithm 22]{b:pjr16}.}
    \State{Set $\widetilde{m}\leftarrow\widetilde{m}+\kappa^{(3)}$, and denote $\{\widetilde{\chi}_j\}_{j=0}^{\widetilde{m}+1}$ as the new temporal resolution.}
    \State{Augment $\mathcal{S}(Z)$, together with $\mathcal{S}(X^{(3)})$, by $\ell_1$ iterative temporal bisections using \citep[Algorithm 22]{b:pjr16}, where $\ell_1$ is sufficient to determine the $j$ for which $[\widetilde{\chi}_j,\widetilde{\chi}_{j+1}]\ni \tau^{(Z)}$.}
    \State{Set $\widetilde{m}\leftarrow\widetilde{m}+\ell_1$, and denote $\{\widetilde{\chi}_j\}_{j=0}^{\widetilde{m}+1}$ as the new temporal resolution.}
    \State{Determine whether $\tau^{(aux)}<\infty$ by application of \algoref{alg:sampling_Z}, Steps 5--13, where $X^{(1)}$ is replaced with $Z$ and $\timerev{}$ is replaced with $X^{(3)}$. Let $\ell_2$ denote the number of additional points inserted.}
    \State{Set $\widetilde{m}\leftarrow\widetilde{m}+\ell_2$, and denote $\{\widetilde{\chi}_j\}_{j=0}^{\widetilde{m}+1}$ as the new temporal resolution.}
    \Until{$\tau^{(aux)}<\infty$}
    \State{Sample $U\sim \texttt{Unif([0,1])}$}
    \If{$U\leq\frac{\mathcal{T}}{\mathcal{T}^{(n)}}$}
    \Return{$\{\mathcal{S}(Z),\mathcal{T}\}$}
    \Else{ \Return $\{\mathcal{S}(Z)^{(n)},\mathcal{T}^{(n)}\}$}
    \EndIf
  \end{algorithmic}
\end{algorithm}


\section{Numerical examples}\label{sec:numerical}


\subsection{An Ornstein--Uhlenbeck process}\label{sec:ou_process}

We consider the computational scaling of the CDB sampler (\algoref{alg:cdb_alg}) as the bridge duration increases (i.e.\ as $T$ increases), contrasted with simulating the bridge directly using path-space rejection sampling: we use the $\varepsilon$-strong simulation approach of \cite{beskos2012varepsilon,b:pjr16}. Recall, the primary motivation of the simple and confluent diffusion bridge approaches is to reduce the simulation cost from exponential to linear in $T$. Our first goal is to illustrate this reduction using the well-studied example of the Ornstein--Uhlenbeck process, defined as the solution to the following SDE
\begin{equation*}
\dd X_t = \theta(\mu-X_t)\dd t + \sigma \dd W_t,\quad X_0=x_0,\quad t\in[0,T].
\end{equation*}
For the experiments below we set $(\theta,\mu,\sigma)=(1,0,1)$ and measure the execution time of the algorithm (in seconds) when applied to the sampling of $0$--$0$ bridges (with $x_0$--$x_T$ bridge denoting a bridge that start in $x_0$ and terminates in $x_T$) over a range of different values of bridge duration.

\begin{figure}[!ht]
\includegraphics[width=0.85\linewidth]{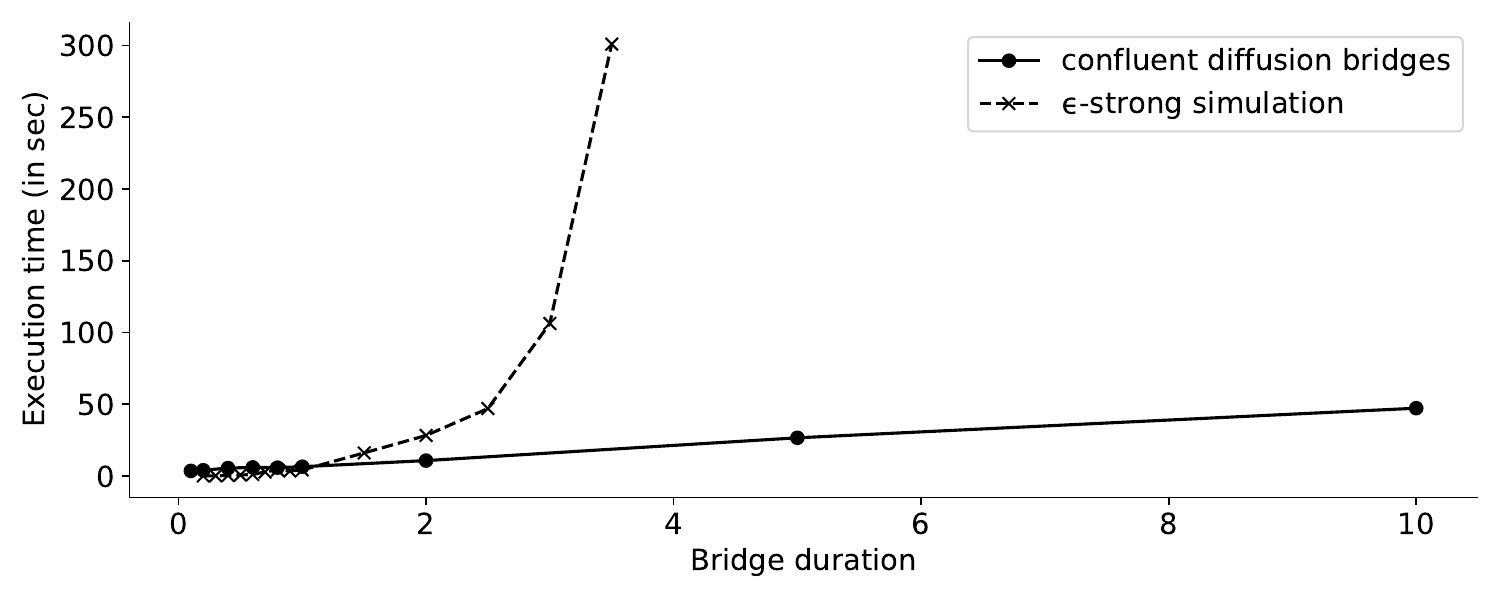}
\caption{Execution time (in seconds) for sampling 100 bridges using confluent diffusion bridges and $\epsilon$-strong simulation} \label{fig:ou_results}
\end{figure}

\figref{fig:ou_results} makes it clear that the computational cost of the confluent diffusion bridges approach scales linearly with $T$, as opposed to the exponential scaling of $\varepsilon$-strong simulation. The latter is more efficient only when the bridge duration is very short. This is in line with our expectation: if the bridge duration is short then path-space rejection sampling will have a large acceptance probability, whereas the confluent diffusion bridge algorithm requires at least three diffusion paths to be simulated. Furthermore, over a short bridge duration it is less likely that $X^{(1)}$ and $\timerev{}$ cross, and that $Z$ and $X^{(3)}$ cross. However, the motivation of this paper is to circumvent the more challenging simulation of diffusion bridges of large duration; in this regime the favourable performance of the confluent diffusion bridge approach is stark. 


\subsection{A Langevin diffusion}\label{sec:langevin_process} 

In this section we consider a far more complicated example inspired by the \emph{Monte Carlo Fusion} approach of \citep{jap:dpr19}. Monte Carlo Fusion is a methodology for unifying distributed analyses and inferences on shared parameters from multiple sources, into a single coherent inference. This is a commonly arising problem in areas such as distributed `big data' problems and privacy settings. The underpinning theory requires the \emph{exact} simulation of (multiple) Langevin diffusion bridge with known invariant distribution. It is natural to consider a confluent diffusion bridge as it does not introduce any approximation errors. 

\begin{figure}[!ht]
\includegraphics[width=0.85\linewidth]{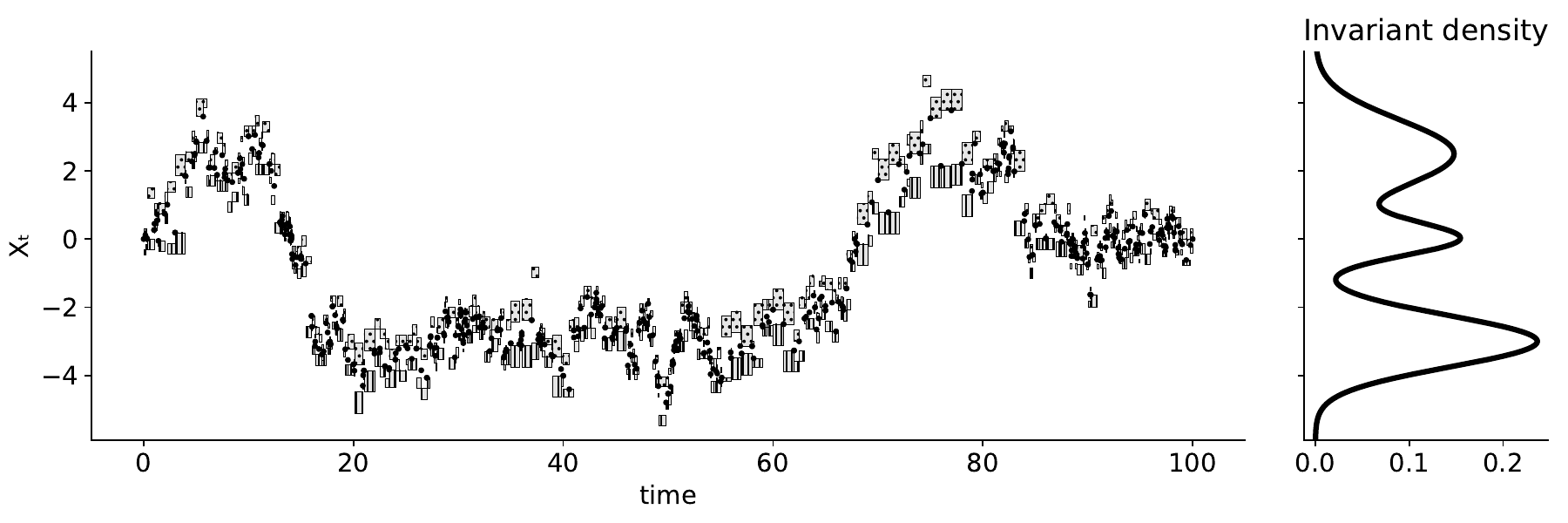}
\caption{\textbf{Left}: An example trajectory of a Langevin diffusion bridge joining $0$ and $0$ over the interval $[0,300]$ and following the diffusion law \eqref{eq:sde_langevin}, simulated using confluent diffusion bridges approach. \textbf{Right}: invariant density of the diffusion \eqref{eq:sde_langevin}.} \label{fig:langevin}
\end{figure}

Consider the diffusion $X$, a solution to the following SDE: 
\begin{equation}\label{eq:sde_langevin}
\dd X_t = \frac{1}{2}\nabla \log f(X_t) \dd t + \dd W_t,\quad X_0 = x_0,\quad t\in[0,T],
\end{equation}
where we set:
\begin{equation}\label{eq:choice_of_f}
f(x):= \sum_{i=1}^3 c_i g(x|\mu_i,\sigma_i),
\end{equation}
with $g(\cdot|\mu,\sigma)$ denoting the probability density function of a Gaussian random variable with mean $\mu$ and standard deviation $\sigma$. The $9$ constants in \eqref{eq:choice_of_f} were set to the following values:
\begin{alignat*}{3}
&c_1 = \frac{10}{27},\qquad && \mu_1 = 2.5,\qquad && \sigma_2 = 1,\\
&c_2 = \frac{5}{27},\qquad && \mu_1 = 0,\qquad && \sigma_2 = 0.5, \\
&c_3 = \frac{12}{27},\qquad && \mu_1 = -3,\qquad && \sigma_2 = 0.75.
\end{alignat*}
Equation \eqref{eq:sde_langevin} defines a Langevin diffusion with invariant density proportional to $f$, which for the particular choice \eqref{eq:choice_of_f} of $f$ becomes a mixture of Gaussian densities. An example trajectory of a $0$--$0$ diffusion bridge, simulated using the confluent diffusion bridges algorithm is given in \figref{fig:langevin}. Note that simulating diffusion bridges of long duration from \eqref{eq:sde_langevin} is a challenging problem: the process can exhibit multimodal behaviour and can yield trajectories that do not resemble paths of Brownian motion.

The numerical simulations conducted had an acceptance rate above $97\%$, suggesting that the proposal bridges are very close to being independent samples from the target diffusion law, and again indicating the confluent diffusion bridge approach can offer substantial computational advantage over competing methodologies. 


\section{Conclusion} \label{sec:conclusion}

In this paper we proposed a novel approach to simulate bridges for ergodic diffusions, based on the combination of three algorithms: $\varepsilon$-strong simulation of Brownian motion, path space rejection sampling, and simple diffusion bridges. The result is a methodology which produces unbiased samples from the target diffusion bridge measure together with bounds on local minima and maxima of the sampled trajectory (as in $\varepsilon$-strong simulation). The resulting algorithm possesses the desirable property that the computational time scales linearly with bridge duration (just as in simple diffusion bridges), while remaining \emph{exact} (just as in path-space rejection sampling). A particularly interesting feature is it is well-suited to bridging over \emph{distant}, rather than \emph{near}, points. That the independence sampler gets better with increasing $T$ contrasts with existing approaches. Indeed, it is also clear that the path space rejection sampler or $\varepsilon$-strong simulation of \citep{beskos2012varepsilon} is still preferable for small $T$. A useful by-product of acquiring local maxima and minima is that it makes it possible to compute, without introducing any approximation errors, various quantities that depend on a path having crossed some pre-specified (random) barriers. Such problems are common for instance in finance, and more precisely in pricing barrier options (see \citep{roberts1997pricing}). 

Our numerical studies in \secref{sec:numerical} verify empirically that we attain the linear-in-time computational scaling. As discussed in \secref{sec:langevin_process} this is particularly useful in practical approaches for unifying distributed analyses and inferences on shared parameters from multiple sources, into a single coherent inference. As such, one fruitful direction of this work would be to embed it within the \emph{fusion} approach of \cite{jap:dpr19}. 

Extending our methodology to a multi-dimensional diffusion bridge setting is a promising direction. Although this seems challenging as we rely on the crossing of unconditioned diffusions, the simple diffusion bridges methodology has in fact been extended to multi-dimensional ergodic diffusions by \citep{jrssb:bfs16}. 

Another clear course for future work is to extend the class of models to include jump diffusion bridges. Again, there has been some work in the context of this for path-space rejection sampling: \citep{mcap:cr11} consider the setting of unconditioned jump diffusions, and \citep{wsc:p15} and \citep{arxiv:glr17} consider conditioned jump diffusions. A confluent diffusion bridge approach would be of particular promise as it could potentially avoid a number of strong (and unnatural) assumptions that are needed in existing rejection sampling algorithms for jump diffusion bridges (in particular, assumptions on the jump intensity).

More broadly, having available diffusion simulation techniques which are linear-in-time between end points opens up new avenues in the developing area of Bayesian inference for diffusions. For instance, it is commonplace when considering inference for discretely observed diffusions to employ data augmentation schemes (or \emph{knots}) \citep{arxiv:mjpr20}, and so this approach could aid insight into the optimal placing of such knots. Another direction is to study whether the methodology could itself be adapted to construct a perfect simulation algorithm for diffusion bridges using coupling-from-the-past (noting that our methodology is itself underpinned by an independence sampler).


\section*{Acknowledgements}
The authors would like to thank Dr Marcin Mider for substantial contributions to the development of this work. They would also like to thank Prof Paul Fearnhead for stimulating discussion on aspects of this paper, and are particularly grateful to Prof Frank van der Meulen whose detailed comments helped identify a methodological problem with an earlier version of this article. PJ, MP, and GOR were supported by The Alan Turing Institute under the EPSRC grant EP/N510129/1. GOR was additionally supported under the EPSRC grants EP/K034154/1, EP/K014463/1, and EP/R018561/1. 


\bibliography{bibliography.bib}

\end{document}